\documentclass[preprint,3p,sort&compress,a4paper]{elsarticle}

\usepackage{graphicx}
\usepackage{subfig}
\usepackage{enumerate} 
\usepackage{amssymb}
\usepackage{amsthm}

\usepackage[pdftex]{hyperref}
\hypersetup{
    colorlinks,%
    citecolor=black,%
    filecolor=black,%
    linkcolor=black,%
    urlcolor=black
}

\begin{document}

\begin{frontmatter}

\title{Linking electromagnetic precursors with earthquake dynamics: an approach based on nonextensive fragment and self-affine asperity models}

 \author[1]{G. Minadakis}
 \author[2]{S. M. Potirakis}
 \author[3]{C. Nomicos}
 \author[4]{K. Eftaxias}
 \address[1]{Department of Electronic and Computer Engineering, Brunel University Uxbridge, Middlesex, UB8 3PH, U.K.}
 \address[2]{Department of Electronics, Technological Educational Institute of Piraeus, 250 Thivon \& P. Ralli, GR-12244, Aigaleo, Athens, Greece}
 \address[3]{Department of Electronics, Technological Educational Institute of Athens, Ag. Spyridonos, Egaleo, GR 12210, Athens, Greece}
 \address[4]{Department of Physics, Section of Solid State Physics, University of Athens, Panepistimiopolis, GR 15784, Zografos, Athens, Greece}

\begin{abstract}
Understanding the earthquake (EQ) preparation process in terms of precursory electromagnetic (EM) emissions has been an evolving field of multi-disciplinary research. EM emissions in a wide frequency spectrum ranging from kHz to MHz are produced by opening cracks, which can be considered as precursors of general fracture. An important feature, observed on both laboratory and geophysical scale, is that the MHz radiation systematically precedes the kHz one. Yet, the link between an individual EM precursor and a distinctive stage of the EQ preparation comprises a crucial open question. A recently proposed two-stage model on preseismic EM activity suggests that the MHz EM emission is due to the fracture of the highly heterogeneous system that surrounds the fault. The finally emerged kHz EM emission is rooted in the final stage of EQ generation, namely, the fracture of entities sustaining the system. In this work we try to further penetrate and elucidate the link of the precursory kHz EM activity with the last stage of EQ generation building on two theoretical models for EQ dynamics. Firstly, the self-affine model states that an EQ is due to the slipping of two rough and rigid fractional Brownian profiles, one over the other, when there is an intersection between them. Secondly, the fragment-asperity model, rooted in a nonextensive Tsallis framework starting from first principles, consists of two rough profiles interacting via fragments filling the gap. In the latter approach, the mechanism of triggering EQ is established through the interaction of the irregularities of the fault planes and the fragments between them. This paper shows that these models of EQ dynamics can be linked with the detected kHz EM emission. In this framework of analysis of preseismic EM activity, we identify sufficient criteria that offer the possibility to discriminate whether a seismic shock is sourced in the fracture of fragments filling the gap between the rough profiles or in the fracture of ``teeth'' distributed across the fractional Brownian profiles that sustain the system.
\end{abstract}


\end{frontmatter}

\section{Introduction}\label{sec:Intro}

The use of basic principles of fracture mechanics for understanding the earthquake (EQ) preparation process is a challenging theme of multi-disciplinary research. Key fundamental questions in strength considerations of materials are: when does it fail? Are there signals that can warn of imminent failure?  It has been shown that fracture induced electromagnetic (EM) physical fields allow real-time monitoring of damage evolution in materials during mechanical loading. EM emissions in a wide frequency spectrum ranging from kHz to MHz are produced by opening cracks, which can be considered as precursors of general fracture \cite{Bahat2005,Frid2003,Fukui2005,Keefe1995,Lacidogna2005,Lolajicek1996,Mavromatou2004,Ogawa1985,Panin2001}. These precursors are detectable both at laboratory \cite{Bahat2005} and geophysical scale \cite{Rabinovitch2001,Gokhberg1995}. An important feature, observed on both scales, is that the MHz radiation systematically precedes the kHz one \cite{Eftaxias2002,Eftaxias2004,Kapiris2004,Contoyiannis2005}. Yet, the link between an individual EM precursor and a distinctive stage of the EQ preparation comprises a crucial open question. 

The present paper draws on a recently proposed two-stage model on preseismic EM activity \cite{Kapiris2004,Contoyiannis2005,Contoyiannis2008,Papadimitriou2008,Eftaxias2006,Eftaxias2007b} suggesting that the MHz EM emission is due to the fracture of the highly heterogeneous system that surrounds the fault. This MHz EM activity can be attributed to phase transition of second order \cite{Contoyiannis2005}, while a Levy walk type mechanism can explain the observed critical state \cite{Contoyiannis2008}. The finally emerged kHz EM emission, from approximately one week up to a few hours before the main shock occurrence, is rooted in the final stage of EQ generation, namely, the fracture of entities sustaining the system \cite{Kapiris2004,Contoyiannis2005,Kalimeri2008,Papadimitriou2008,Eftaxias2009}. 

In this work we try to further penetrate and elucidate the link of the precursory kHz EM activity with the last stage of EQ generation building on two models for EQ dynamics. First, De Rubeis et al., (1996) \cite{De-Rubeis1996} and Hallgass et al., (1997) \cite{Hallgass1997} introduced a self-affine model for fault dynamics by means of the slipping of two rough and rigid Brownian profiles one over the other. In this self-affine asperity scheme, an individual EQ occurs when there is intersection between the two profiles. The second model concerns a fragment-asperity model for EQ dynamics which has been recently introduced by Solotongo-Costa and Posadas (SCP) \cite{Sotolongo2004}. This model is rooted in a nonextensive Tsallis framework starting from first principles, and draws on a scheme of two rough profiles interacting via fragments filling the gap. 

Herein, we argue that the aforementioned two models for EQ dynamics are identified in two qualitatively different epochs of kHz EM precursory activity: 
\begin{enumerate}[(i)]
\item {the first epoch of the kHz EM precursor reflects the breakage of the fragments filling the gap between the two profiles.}
\item {the second epoch is emitted during the fracture of large and strong entities (''teeth'') which are distributed along these two rough profiles of the fault.}
\end{enumerate}

The structure of this paper is organized as follows: In section 2, we describe the aforementioned self-affine and fragment-asperity model for EQ dynamics. In Sec. 3 we set the context for the proposed model drawing on recent studies of kHz EM precursory activity in terms of complexity-organization. In Sec. 4 we analyze the kHz EM recordings by means of the Hurst's rescaled range analysis, distinguishing two emitted epochs that reflect on two different fracture regimes. In Sec. 5 we give the definition of the ``electromagnetic earthquake'' (EM-EQ) and we examine the nonextensive behaviour of the kHz EM activity focusing on the link between the two identified epochs associated with the last stage of EQ preparation process. Sec. 6 draws on the footprints of fractal Brownian motion (fbm) profile and roughness of surface fracture related with the two epochs of kHz EM activity. In the latter part of that section we provide argumentation for the association of the second epoch with the self-affine model. Finally we summarize the key findings that support the proposed approach in view of further research considerations. 

\section{Overview of two models for earthquake dynamics}\label{sec:models}

\subsection{A self-affine asperity model for earthquake dynamics: experimental and theoretical evidence}
\label{self_affine_model}

De Rubeis et al., (1996) \cite{De-Rubeis1996} and Hallgass et al., (1997) \cite{Hallgass1997} have worked on a model for regional fault dynamics by means of the slipping of two rough and rigid Brownian profiles one over the other. In this scheme, an individual EQ occurs when there is intersection between the two profiles. The authors assumed that the energy released is proportional to the extension of the overlap between the two asperities in contact. We briefly describe the rules of the model as defined by De Rubeis et al., (1996) \cite{De-Rubeis1996}: 

\begin{enumerate}[(i)]
\item {The initial condition is obtained by putting two rigid profiles in contact in the point where the height difference is minimal.}
\item {The successive evolution is obtained by drifting a profile in a parallel way with respect to another.} 
\item {An intersection represents a single seismic event and starts with the collision of two asperities of the profiles.} 
\end{enumerate}

This model exhibits a good interpretation of the seismicity generated in a large geographic area usually identified as ``seismic region'', covering many geological faults, in a global sense. Ample experimental and theoretical evidence support the above mentioned scheme: 

\begin{enumerate}[(i)]
\item {Kinematic or dynamic source inversions of EQs suggest that the final slip (or the stress drop) has a heterogeneous spatial distribution over the fault (see among others \citet{Gusev1992,Bouchon1997,Peyrat2001}).}

\item {A study on Power spectrum analysis of the fault surface suggests that heterogeneities are observed over a large range of scale lengths [see Power et al., (1987), \cite{Power1987} in particular  [Fig. 4]].}

\item {Investigators of the EQ dynamics have already pointed out that the fracture mechanics of the stressed crust of the earth forms self-similar fault patterns, with well-defined fractal dimensionalities \cite{Kagan1980,Sahimi1993,Barriere1991}.}

\item {Following the observations of the self-similarity in various length scales in the roughness of the fractured solid surfaces, Chakrabarti et al. (1999) \cite{Chakrabarti1999} have proposed that the contact area distribution between two fractal surfaces follows a unique power law.}

\item {Huang and Turcotte, (1988) \cite{Huang1988} pointed out that natural rock surfaces can be represented by fractional Brownian surfaces over a wide range.}
\end{enumerate}

Interestingly, Hallgass et al. (1997) \cite{Hallgass1997} have emphasized that ``what is lacking is the description of what happened locally, i.e., as a consequence of a single event''. In the following, we attempt to show that the statistics of regional seismicity could be merely a macroscopic reflection of the preparation process of a single EQ.

\subsection{A fragment-asperity model for earthquake dynamics coming from nonextensive statistical mechanics}
\label{fragment_asp_model}

In seismology, the scaling relation between magnitude and the number of EQs is given by the Gutenberg-Richter (G-R) relationship:

\begin{equation}
\log N(>M)=\alpha -bm
\label{G-R Formula}
\end{equation}

where, $N(>M)$ is the cumulative number of EQs with a magnitude greater than $M$ occurring in a specified area and time and $b$ and $\alpha$ are constants. It should be noted that the G-R empirical relation is not related with general physical principles.

It is well-known that the Boltzmann-Gibbs statistical mechanics works best in dealing with systems composed of either independent subsystems or interacting via short-range forces, and whose subsystems can access all the available phase space. For systems exhibiting long-range correlations, memory, or fractal properties, nonextensive statistical mechanics becomes the most appropriate mathematical framework \cite{Tsallis1988,Tsallis2009}. A central property of the EQ preparation process is the occurrence of coherent large-scale collective with a very rich structure, resulting from the repeated nonlinear interactions among its constituents. Consequently, the nonextensive statistical mechanics is an appropriate arena to investigate the last stage of EQ preparation process.

As concerns the material between the fault planes, Herrmann and Roux (1990) \cite{Herrmann1990} studied the phenomenon of fault slipping from a geometric viewpoint, offering an idealized representation of the fragmented core of a fault (gouge). They have presented the gouge as a self-medium formed by circular disk-shaped pieces which act like bearings filling the space between two planes \cite{ Herrmann1990a}. In that direction, Sotolongo-Costa and Posadas (SCP) \cite{Sotolongo2004} have developed a model for EQ dynamics coming from a non-extensive Tsallis formalism, starting from fundamental principles. In particular, the theoretical components of the SCP model read as follows:

\begin{enumerate}[(i)]

\item {The mechanism of relative displacement of fault plates is the main cause of EQs.}

\item {The surfaces of the tectonic plates are irregular. The space between fault planes is filled with the residue of the breakage of the tectonic plates, from where the faults have originated.}

\item {The fragments are very diverse and have irregular shapes. The motion of the fault planes can be hindered not only by the overlapping of two irregularities/asperities (teeth) of the profiles, but also by the eventual relative position of several fragments. Thus, the mechanism of triggering EQs is established through the combination of the irregularities of the fault planes on one hand and the fragments between them on the other hand.}

\item {In this regard, the authors studied the influence of the size distribution of fragments on the energy distribution of EQs.}

\end{enumerate}

We note that the fragments size distribution function comes from a nonextensive Tsallis formulation, starting from first principles, i.e., a nonextensive formulation of the maximum entropy principle. Englaman et al. \cite{Englman1987} showed that the standard Boltzmann-Gibbs formalism, although useful, cannot account for an important feature of the fragmentation process, i.e., the presence of scaling in the size distribution of fragments, which is one of the main ingredients of the SCP approach. Interestingly, the latter nonextensive approach leads to a G-R type law for the magnitude distribution of EQs (see Eq (8) in Ref. \cite{Sotolongo2004}). 

The fragment-asperity (SCP) model has been recently revised by Silva et al. \cite{Silva2006} with two crucial updates. They use a different definition of the mean values in the context of Tsallis nonextensive statistics as proposed by Abe and Bagci \cite{Abe2005}. Moreover, Silva et al. have introduced a new scale law, $\varepsilon \propto r^3$, between the released energy $\varepsilon$ and the size $r$ of fragments. The new scale proposed by Silva et al. \cite{Silva2006} is in full agreement with the standard theory of rupture, namely, the well-known seismic moment scaling with rupture length (see Ref. \cite{Vilar2007} for details). Finally, their approach leads to the following G-R type law for the magnitude distribution of EQs:

\begin{eqnarray}
\label{eq:Silva}
\log \left[N\left(>M\right)\right]=\log N+\left({2-q\over 1-q} \right)\log \left[1-\left({1-q\over 2-q} \right)\left({10^{2M} \over a^{2/3} } \right)\right] 
\end{eqnarray}

\noindent where, $N$ is the total number of EQs, $N(>M)$ the number of EQs with magnitude larger than $M$, and $M\approx \log (\varepsilon)$. The parameter $\alpha$ is the constant of proportionality between the EQ energy, $\varepsilon$, and the size of fragment. The authors mentioned that this is not a trivial result, and incorporates the characteristics of nonextensivity into the distribution of EQs by magnitude. 

The entropic index $q$ characterizes the degree of non-extensivity of the system reflected in the following pseudo-additivity rule: 

\begin{equation}
S_{q}(A+B)=S_{q}(A)+S_{q}(B)+(1-q)S_{q}(A)S_{q}(B)
\end{equation}

where,
\begin{equation}
S_{q}=k\frac{1}{q-1}\left(1-\sum_{i=1}^{W}p_{i}^{q}\right)
\end{equation}

is the Tsallis entropy, $p_{i}$ are the probabilities associated with the microscopic configurations, $W$ is their total number, $q$ is the nonextensive entropic parameter, and $k$ is Boltzmann's constant. 

Importantly, from most geological analysis performed so far, values of $q\sim 1.6-1.8$ seem to be universal, in the sense that different data sets from different regions of the globe indicate a value for the nonextensive parameter lying in this interval \cite{Sotolongo2004,Silva2006,Vilar2007,Telesca2010,Telesca2011}.

\section {Linking earthquake dynamics and kHz electromagnetic activity: setting the context for the proposed approach}
\label{sec:proposal}

\begin{figure}[h]
  	\centering
	 \includegraphics[width=1\textwidth]{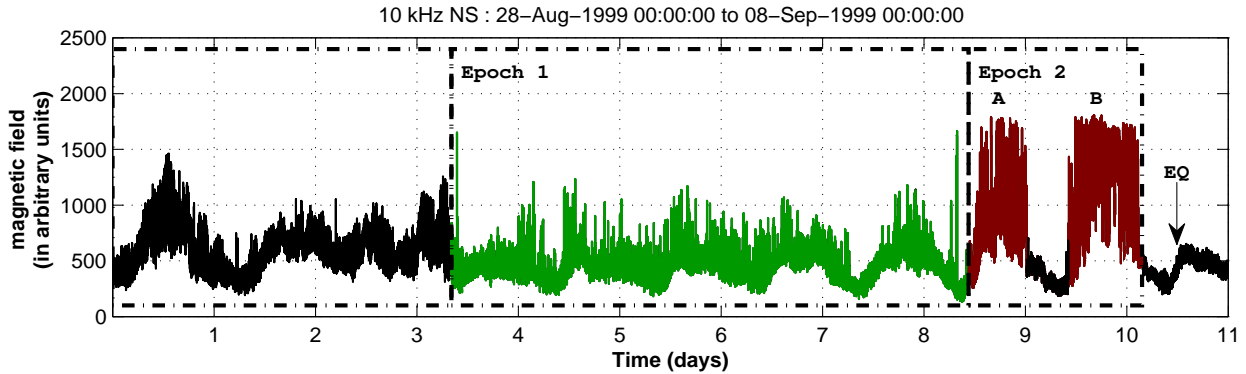}                 
	\caption{View of the preseismic EM emission (green and red) recorded by the 10 kHz NS magnetic field sensor. The vertical arrow indicates the time of the Athens EQ occurrence.}
\label{fig:athsig}
\end{figure}

Our analysis focuses on a well documented \cite{Kapiris2004,Contoyiannis2005,Contoyiannis2008,Papadimitriou2008} kHz EM precursor (depicted in Fig. \ref{fig:athsig} with green and red color) associated with the Athens EQ that occurred on September 7, 1999 at 11:56 (GMT) with magnitude $M_W=5.9$. In this paper, the preseismic magnetic field recorded by the 10 kHz North-South (NS) sensor is studied. 

Drawing from recent studies, multidisciplinary analysis applied on recordings from the 10 kHz East-West (EW) sensor, has shown that the black part in Fig. \ref{fig:athsig} refers to the EM background activity (noise) which has been characterized by a low organization (or high complexity) \cite{Karamanos2006,Karamanos2005}. The first part of the precursor (epoch 1) has been characterized by the appearance of a population of EM events sparsely distributed in time with noteworthy higher order of organization in comparison to that of the noise. The strong bursts A and B in epoch 2 are characterized by a population of EM events of significantly higher organization in comparison to that of epoch 1 which are densely distributed in time. 

\begin{figure}[h]
  \centering
	\includegraphics[width=1\textwidth]{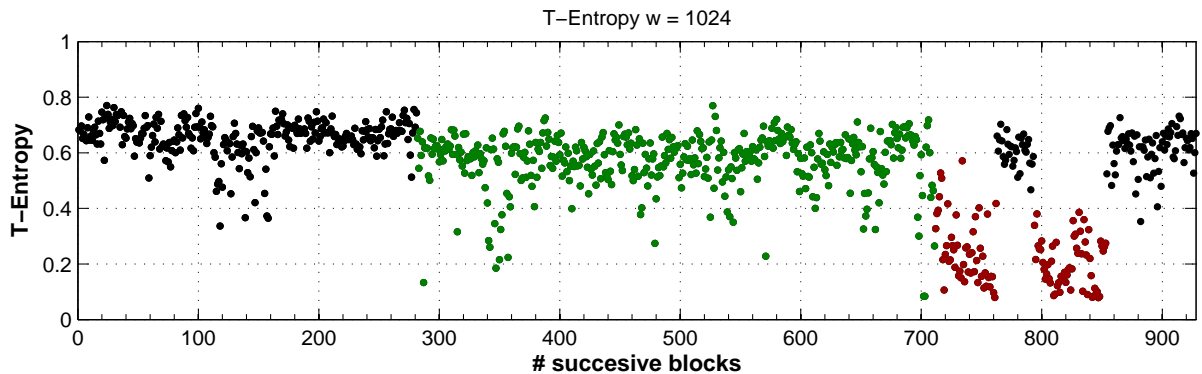}             
\caption{Temporal evolution of T-Entropy using sequential blocks of 1024 samples each.}
\label{fig:T-Entropy}
\end{figure}

The same results are further verified here by the analysis of the recordings of the 10 kHz NS sensor. Characteristically in Fig. \ref{fig:T-Entropy} we present the degree of organization between the two distinctive epochs namely 1 and 2, by means of a robust grammar-based complexity/information technique, the T-Entropy. The analysis has been applied in terms of symbolic dynamics by using sequential successive windows of 1024 samples each. We recall that T-entropy is based on the intellectual economy one makes when rewriting a string according to some rule. The estimation of T-Entropy has been presented in \citet{Karamanos2006,Eftaxias2009}. 

\begin{figure}[h]
  \centering
	\subfloat[]{\label{fig:model1} \includegraphics[width=0.4\textwidth]{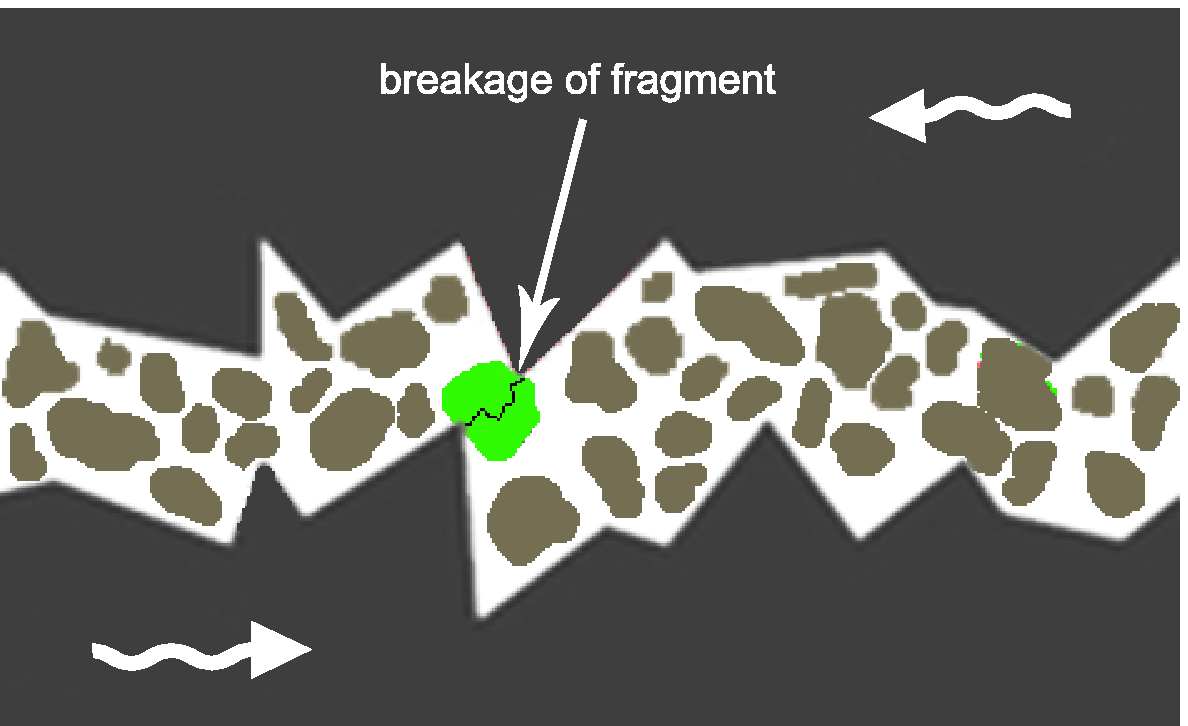}}     
	\quad
	\subfloat[]{\label{fig:model2} \includegraphics[width=0.4\textwidth]{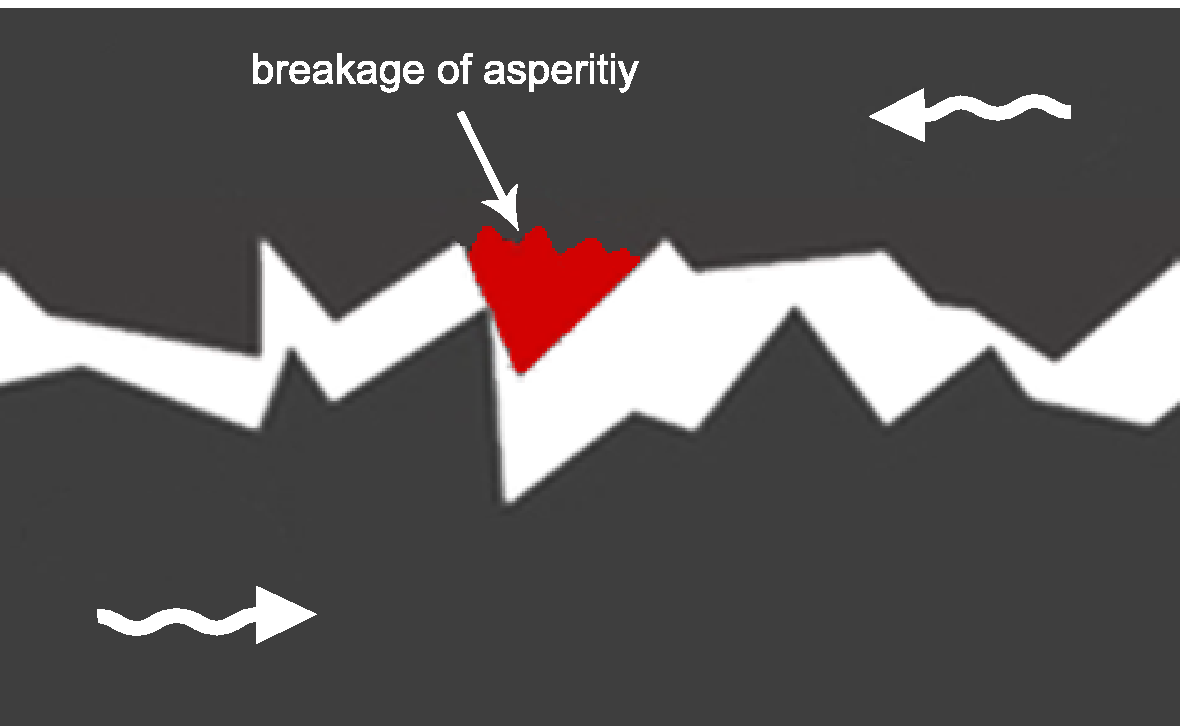}}                    
\caption{(a) An illustration of the fragment-asperity model. EM fluctuations are emitted during the fracture of fragments. (b) An illustration of the self-affine model. EM fluctuations are emitted during the fracture of strong and large ``teeth''.}
\label{fig:model}
\end{figure}

The aforementioned results intuitively lead to the suspicion that the underlying fracture mechanisms are different in epoch 1 and epoch 2, characterized by a negative and positive-feedback mechanism, correspondingly. Against the theoretical background set in the previous section, herein we describe our proposed approach that aims to prove that the two models for EQ dynamics (described in Sec. \ref{sec:models}) are identified in two qualitatively different epochs of the kHz EM precursory activity:

\begin{enumerate}[(i)]
\item {The first epoch (Fig. \ref{fig:athsig}, green color) of the kHz EM precursor may reflect the breakage of the fragments filling the gap between the two profiles that sustain the fault (Fig. \ref{fig:model1}). In this scheme, EM fluctuations are emitted during the fracture of fragments.} 

\item {The second epoch (Fig. \ref{fig:athsig}, red color) is possibly emitted during the fracture of large and strong entities (``teeth'') which are distributed along two rough and rigid fractal Brownian profiles one over the other (Fig. \ref{fig:model2}). In this scheme, EM fluctuations are emitted during the fracture of strong and large ``teeth''.}
\end{enumerate}

In the next section we intend to verify that the two precursory epochs are rooted in different fracture regimes. For this purpose we first use the ``Rescaled Range Analysis (R/S)'' \cite{Hurst1965}.

\section{Analysis of presismic kHz EM emission in terms of the Hurst exponent}
\label{sec:RSAnalysis}

After the works of Hurst et al. \cite{Hurst1965}, Mandelbrot and Wallis \cite{Mandelbrot1968} and Feder \cite{Feder1988}, Hurst's rescaled analysis has been used as a method to detect correlations in time series. There are two factors used in the R/S analysis: first, the range R, which is the difference between the maximum and minimum amounts of accumulated departure of the time series from the mean over a time span $\tau$, and second, the standard deviation $S$ calculated over the time span $\tau$. The so-called rescaled range is exactly the ratio of $R$ and $S$. From a variety of time series of natural phenomena, it is concluded that the ratio R/S is very well described by the following empirical relation:

\begin{equation}
R/S=(\tau/2)^H
\label{Hurst}
\end{equation}

The correlation of the past and the future in the observational time series can be described by the $H$. For $H=0.5$, there is an independent random process, with no correlations among samples. For $H>0.5$, the sequence is characterized by a persistent behaviour, which means that the increasing or decreasing trend may followed by the same positive sign. For $H<0.5$, the sequence is characterized by the anti-persistent behaviour, which means that an increasing or decreasing trend is more likely to be reversed implying a negative feedback mechanism.

\begin{figure}[h]
  \centering
	\includegraphics[width=1.0\textwidth]{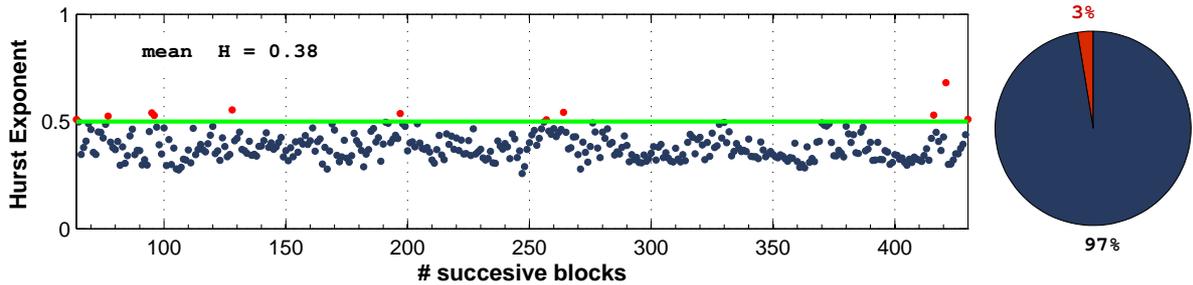}  	   
\caption{Temporal evolution of Hurst exponent $(H)$ for epoch 1, deriving from fixed sequential windows of 1024 samples each. The green line is the threshold of the transition from antipersistent to persistent behavior. The pie shows that 97\% of the blocks calculated, are antipersistent.}
\label{fig:RSepoch1}
\end{figure}

{\it Focusing on the first epoch}, we study the temporal evolution of the $H$ exponent by applying a sequence of successive fixed time windows of 1024 samples each. The results obtained from Fig. \ref{fig:RSepoch1}, reveal that the first epoch is really characterized by anti-persistent behaviour ($0<H<0.5$) with a mean $\bar{H}=0.38$. 

\begin{figure}[h]
  \centering
	\subfloat[]{\label{fig:RS2a}\includegraphics[width=0.5\textwidth]{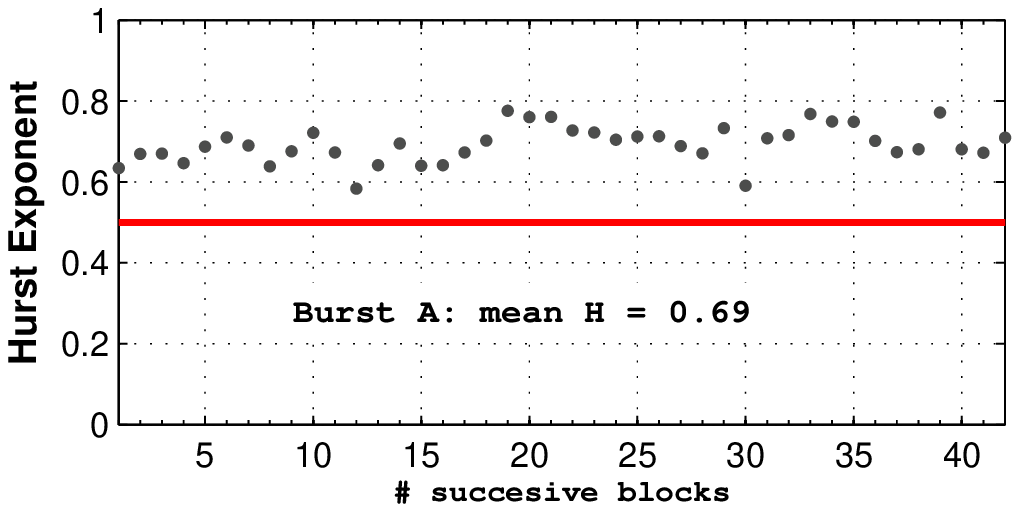}}     
  	\subfloat[]{\label{fig:RS2b}\includegraphics[width=0.5\textwidth]{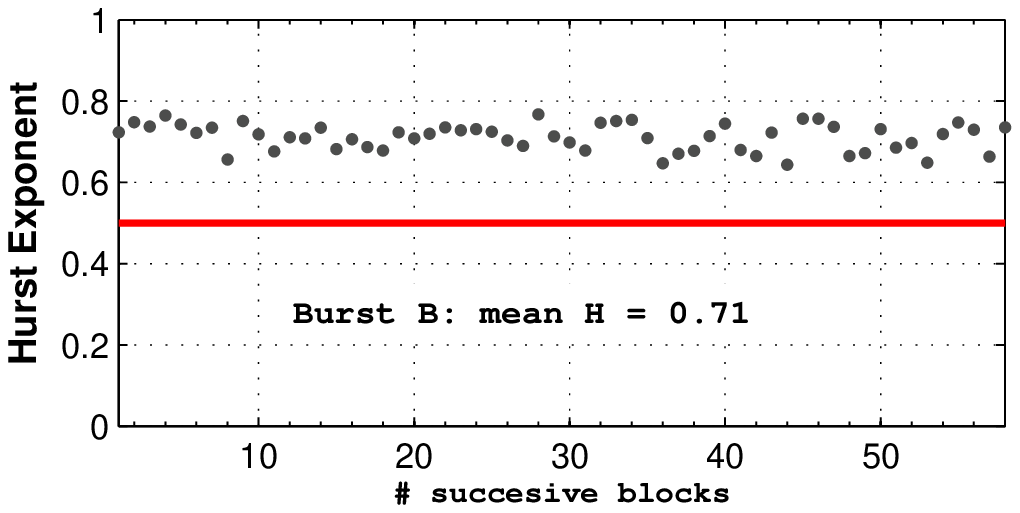}}          
\caption{Temporal evolution of Hurst exponent $(H)$ for the two strong EM bursts contained in epoch 2.}
\label{fig:RSepoch2}
\end{figure}

{\it Focusing on the second epoch}, the same windowing method is used as that of epoch 1. Figs \ref{fig:RS2a} and \ref{fig:RS2b} depict the temporal evolution of $H$ exponent for the two strong EM bursts contained in epoch 2. Results obtained from these figures reveal that both the two EM bursts are characterized by a persistent ($0.5 < H < 1$) behaviour, with a mean $\bar{H}=0.69$ and $\bar{H}=0.71$ respectively.

\begin{figure}[h]
  \centering
	\subfloat[]{\label{fig:RSO2a}\includegraphics[width=0.4\textwidth]{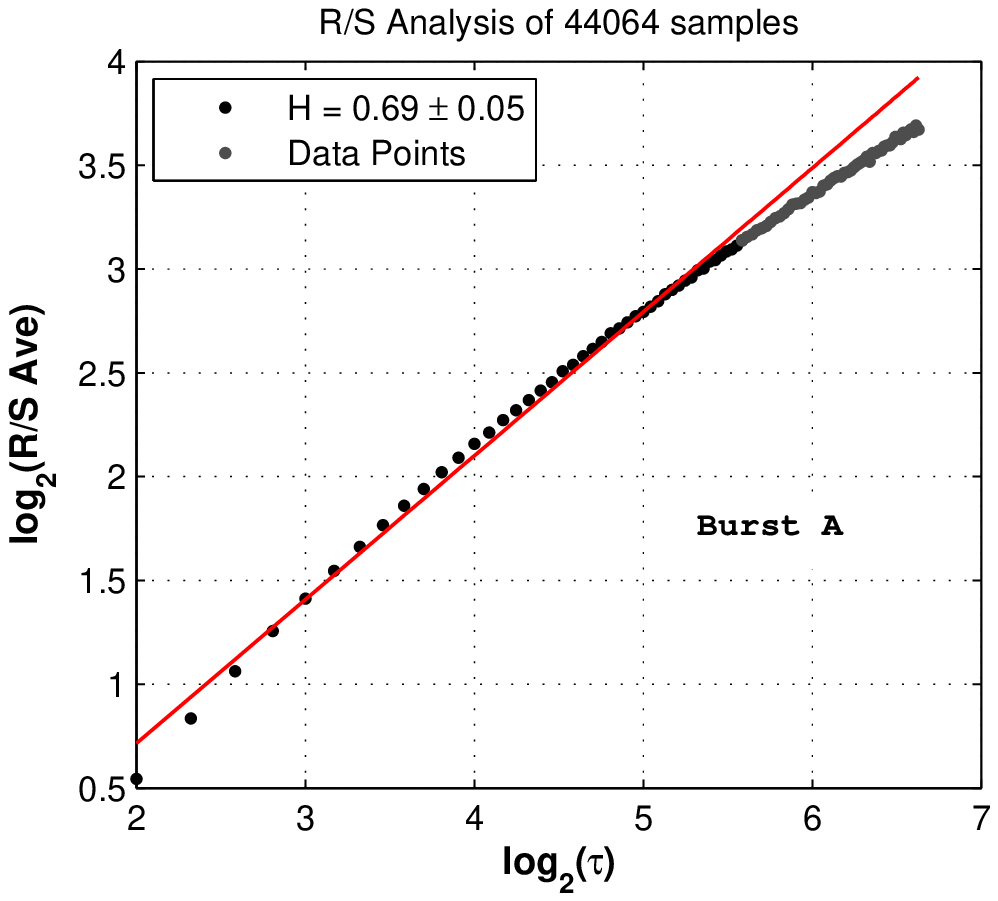}}     
  	\subfloat[]{\label{fig:RSO2b}\includegraphics[width=0.4\textwidth]{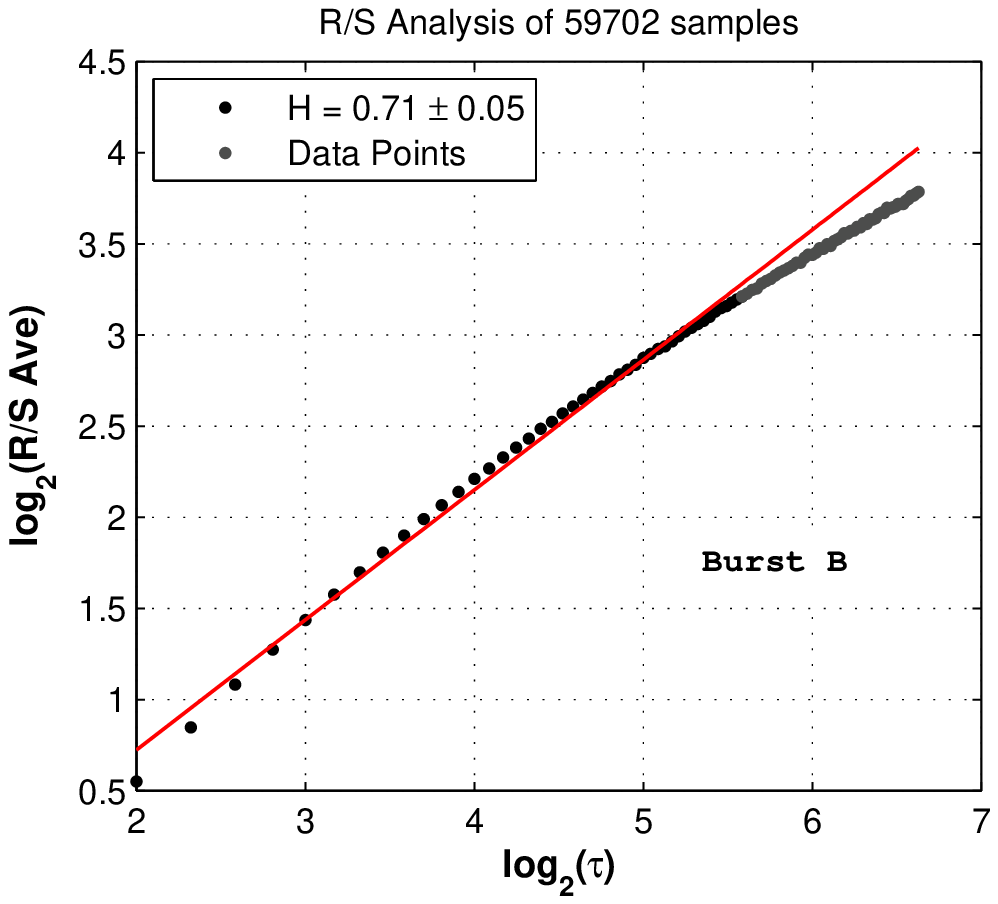}}          
\caption{Hurst exponent estimated in terms of R/S analysis for the whole part of each one of the two strong EM bursts contained in epoch 2.}
\label{fig:RSFepoch2}
\end{figure}

To further support the persistency of the two strong EM bursts we applied the R/S method to the whole part of each one. Figs \ref{fig:RSO2a} and \ref{fig:RSO2b} depict the slopes deriving from the linear regression fitting of R/S method with $H=0.69 \pm 0.05$ and $H=0.71 \pm 0.05$ for the first and second EM bursts, respectively. The derived $H$ exponents along with those of the windowing method strongly reveal that both EM bursts are characterized by an onset of persistency. 

Summarizing, the results of R/S analysis strongly support the hypothesis that the two preseismic epochs are rooted in different regimes characterized by negative and positive feedback correspondingly. To further investigate the possible differentiation of the two epochs we turn to the nonextensive model as described in section \ref{fragment_asp_model}.

\section{Nonextensive approach of the two distinct epochs of kHz EM activity}
\label{sec:scpsilva}

The analysis presented in this section first deals with the examination of whether the nonextensive formula \ref{eq:Silva} also describes the sequence of ``electromagnetic earthquakes'' (EM-EQs) included in the recorded EM precursor. Should this case exist, we go on to further examine the variations of the non-extensive parameter $q$ and energy density $\alpha$ for different thresholds of magnitudes of the detected sequence of the EM-EQs. 

{\it The definition of electromagnetic earthquake}: Herein, we regard as amplitude $A$ of a candidate ``fracto-electromagnetic fluctuation'' the difference $A_{f_{EM}}(t_{i})=A(t_{i})-A_{noise}$, where $A_{noise}$ is the background (noise) level of the EM time series. We consider that a sequence of $k$ successively emerged ``fracto-electromagnetic fluctuations'' $A_{f_{EM}}(t_{i})$, $i=1,\ldots,k$ represents the damage of a fragment or a ``tooth'' in the irregular surfaces of fault. We refer to this as an ``EM-EQ''. Since the squared amplitude of the fracto-EM emissions is proportional to their energy $\varepsilon_{EM}$, the magnitude $M$ of the candidate EM-EQ is given by the relation: 

\begin{equation}
M=\log {\varepsilon_{EM}}\sim\log \left( \sum{{{\left[ {{A}_{f_{EM}}}({{t}_{i}}) \right]}^{2}}} \right)
\end{equation}

In order to achieve an efficient and accurate fit for the estimation of the nonextensive parameters $q$ and $\alpha$, the Levenberg-Marquardt (LM) method \cite{Levenberg1944,Marquardt1963,Gallant1975,Bates1988,Press1992} was applied. The LM method is used to solve nonlinear least squares problems when fitting a parameterized function to a set of given data points. Fitting is achieved by minimizing the sum of the squares of the errors between the data points and the function.

\begin{figure}[h]
  \centering
	\subfloat[]{\label{fig:Silva1}\includegraphics[width=0.4\textwidth]{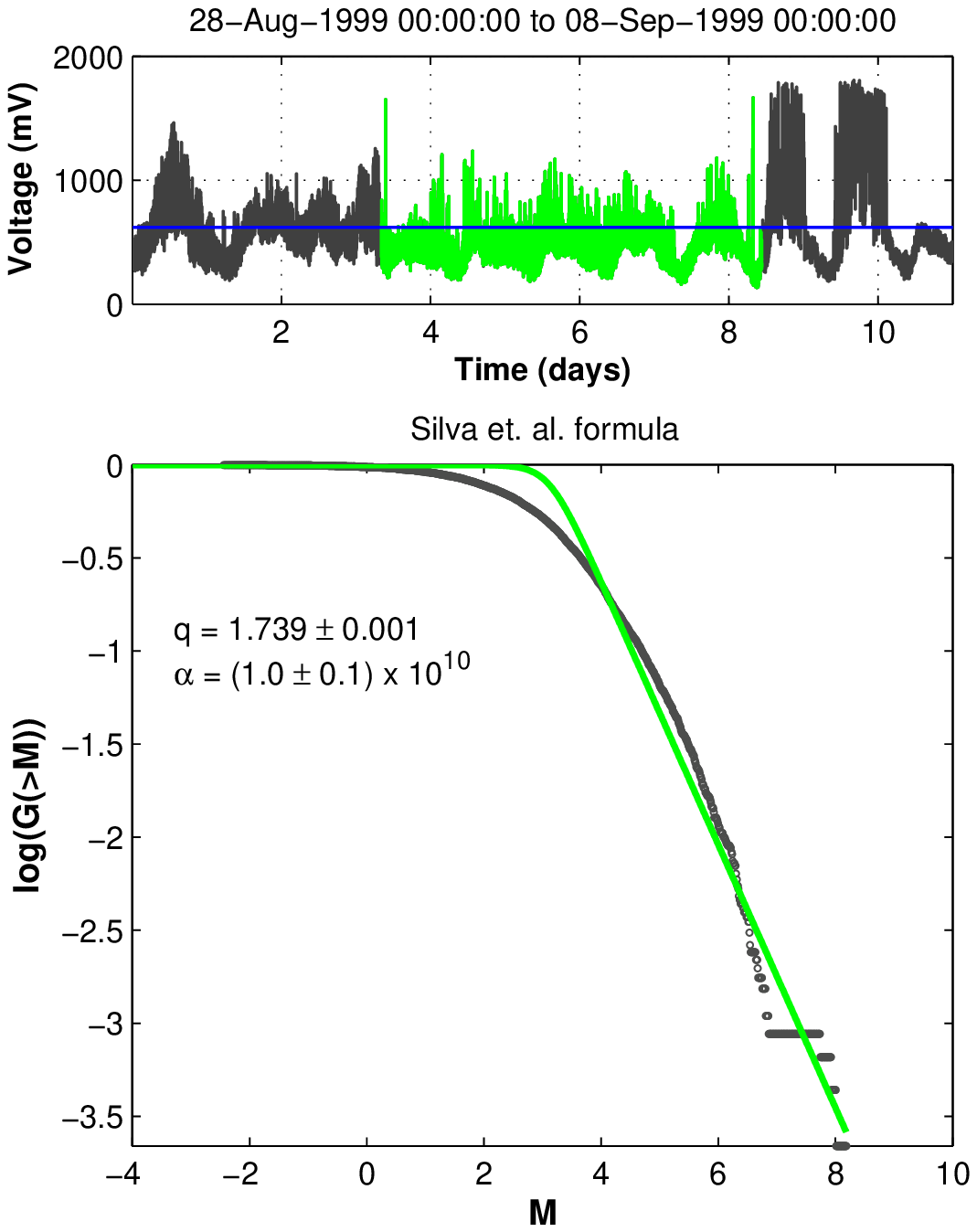}}
	\quad
  	\subfloat[]{\label{fig:Silva2}\includegraphics[width=0.4\textwidth]{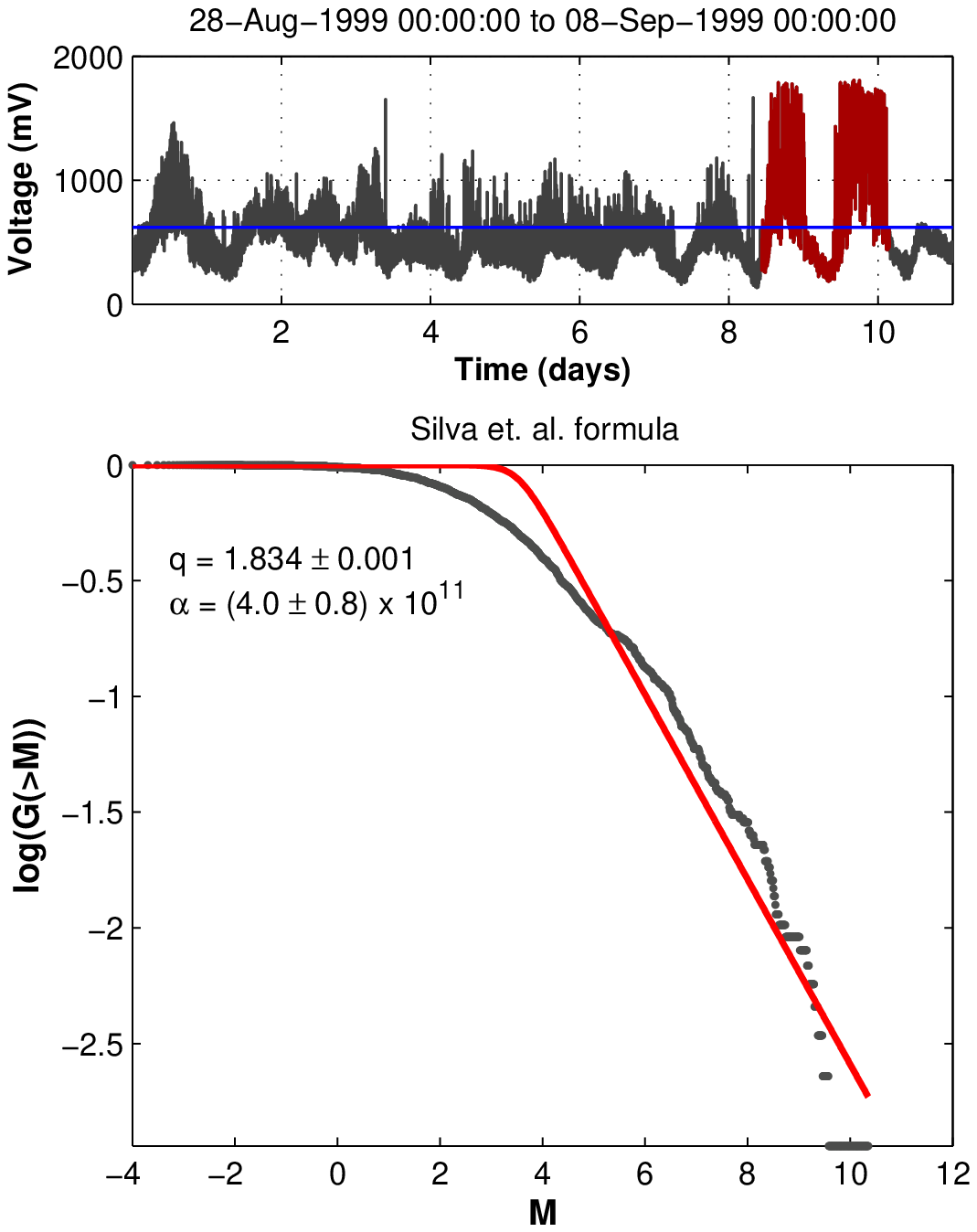}}    
\caption{We use Eq. \ref{eq:Silva} to calculate the relative cumulative number of EM-EQs, $G(> M)$, for epochs 1 and 2 respectively. For the calculation of EM-EQs we used as ${{A}_{noise}}$ the threshold of $620mV$}
\label{fig:Silva}
\end{figure}

Figs. \ref{fig:Silva1} and \ref{fig:Silva2} show that Eq. \ref{eq:Silva} along with the use of LM method provides an excellent fit to the kHz EM-EQs included in the epochs 1 and 2, respectively, incorporating the characteristics of nonextensivity statistics into the detected kHz EM precursor.
Herein, $N$ is the total number of the detected EM-EQs, $G(>M)=N(M>)/N$ the normalized cumulative number of EM-EQs with magnitude larger than $M$, and $\alpha$ the constant of proportionality between the EM energy released and the size of fragment \cite{Sotolongo2004,Silva2006}. The best-fit parameters for this analysis are given by $q = 1.739 \pm 0.001$ for epoch 1 and $q = 1.834 \pm 0.001$ for epoch 2 respectively. These values evidently reveal that both epochs 1 and 2 are characterized by high non-extensivity. 

It is very interesting to observe the similarity in the $q$ values associated with Eq. \ref{eq:Silva} for all the catalogs of EQs used ($q\sim 1.6-1.8$, see Sec. \ref{sec:scpsilva}), as well as for all the precursory sequences of EM-EQs under study. The observed similarity in the $q$ values also indicates that the activation of a single EQ (fault) could be considered as a reduced self-affine image of the whole regional seismicity. These results are consistent with the self-affine nature of fracture and faulting processes \cite{Mandelbrot1985}.

We clarify that the parameter $q$ itself is not a measure of the complexity-organization of the system but only a measure for the degree of non-extensivity. The dynamic changes of the complexity of the system can be quantified by applying the Tsallis entropy equation ($S_{q}$) to the time variation of the signal with a given nonextensive parameter $q$ \cite{Kalimeri2008}. In that case, lower $S_{q}$ values characterize the sequential portions of the signal with lower complexity (or higher organization). 

\begin{figure}[h]
  \centering
	\subfloat[epoch 1]{\label{fig:tempvartsallis1}\includegraphics[width=0.5\textwidth]{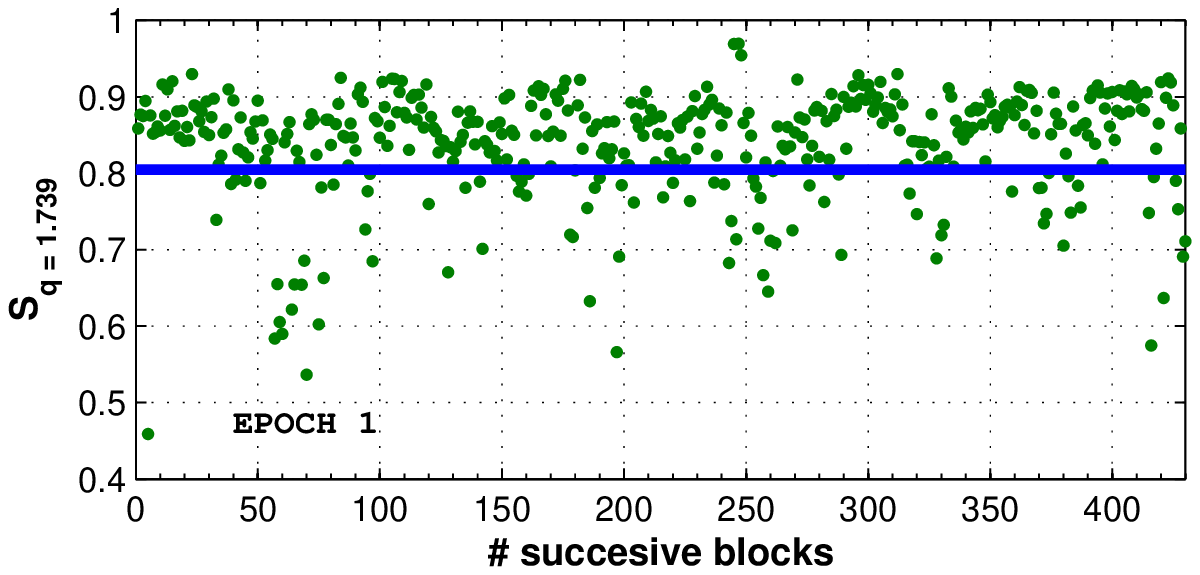}}    
  	\subfloat[epoch 2]{\label{fig:tempvartsallis2}\includegraphics[width=0.5\textwidth]{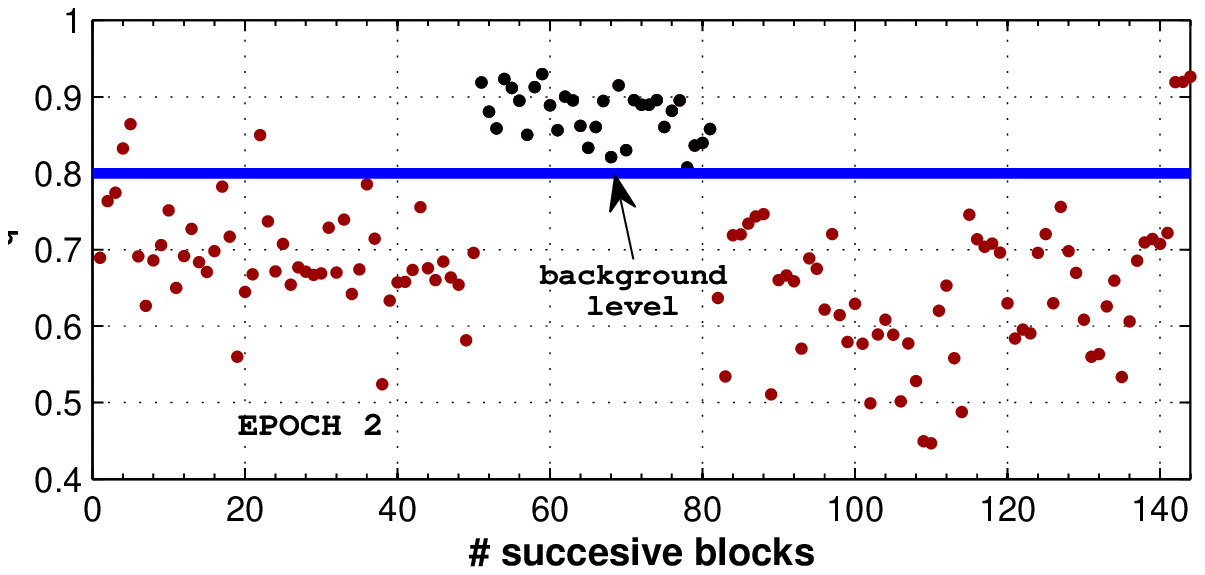}}           
\caption{Temporal evolution of Tsallis entropy for epochs 1 and 2 respectively. The blue line depicts the entropy level that refers to the background EM activity.}
\label{fig:tempvartsallis}
\end{figure}

Figs. \ref{fig:tempvartsallis1} and \ref{fig:tempvartsallis2} depict the temporal variation of the Tsalis entropy for the epochs 1 and 2 respectively. The analysis has been applied in terms of symbolic dynamics by using sequential successive windows of 1024 samples each. For each epoch, we use the estimated parameters $q$ as derived from Eq. \ref{eq:Silva}: $q=1.739$ for epoch 1 and $q=1.834$ for epoch 2. A detailed analysis of Tsallis entropy is presented in \cite{Kalimeri2008}. The results of this statistical analysis are consistent with those obtained from the T-Entropy in section \ref{sec:proposal}. As shown, epoch 2 is characterized by a high organized population of EM events which are densely distributed in time in contrast to epoch 1, which is characterized by a population of lower organization sparsely distributed in time. Both epochs (1,2) have higher organization in respect to that of the background noise.

\begin{figure}[h]
  \centering
	\subfloat[]{\label{fig:tempvar1}\includegraphics[width=0.5\textwidth]{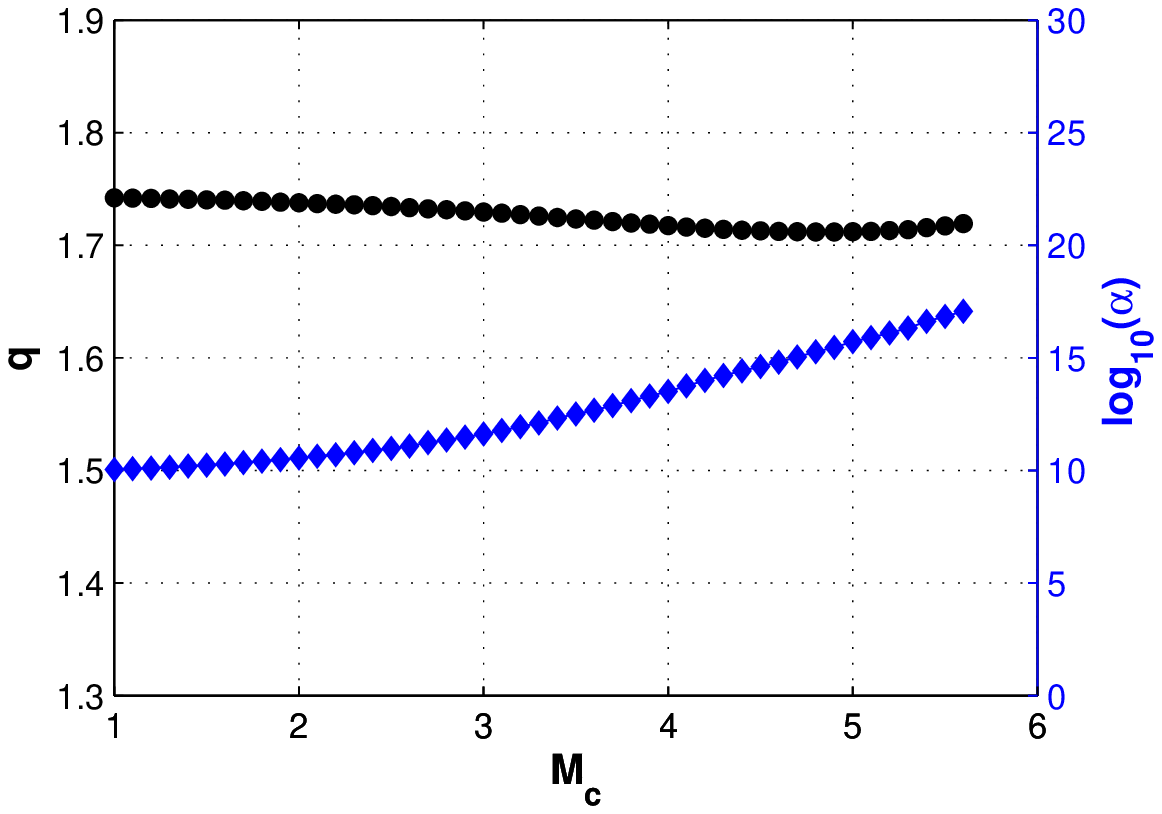}}     
  	\subfloat[]{\label{fig:tempvar2}\includegraphics[width=0.5\textwidth]{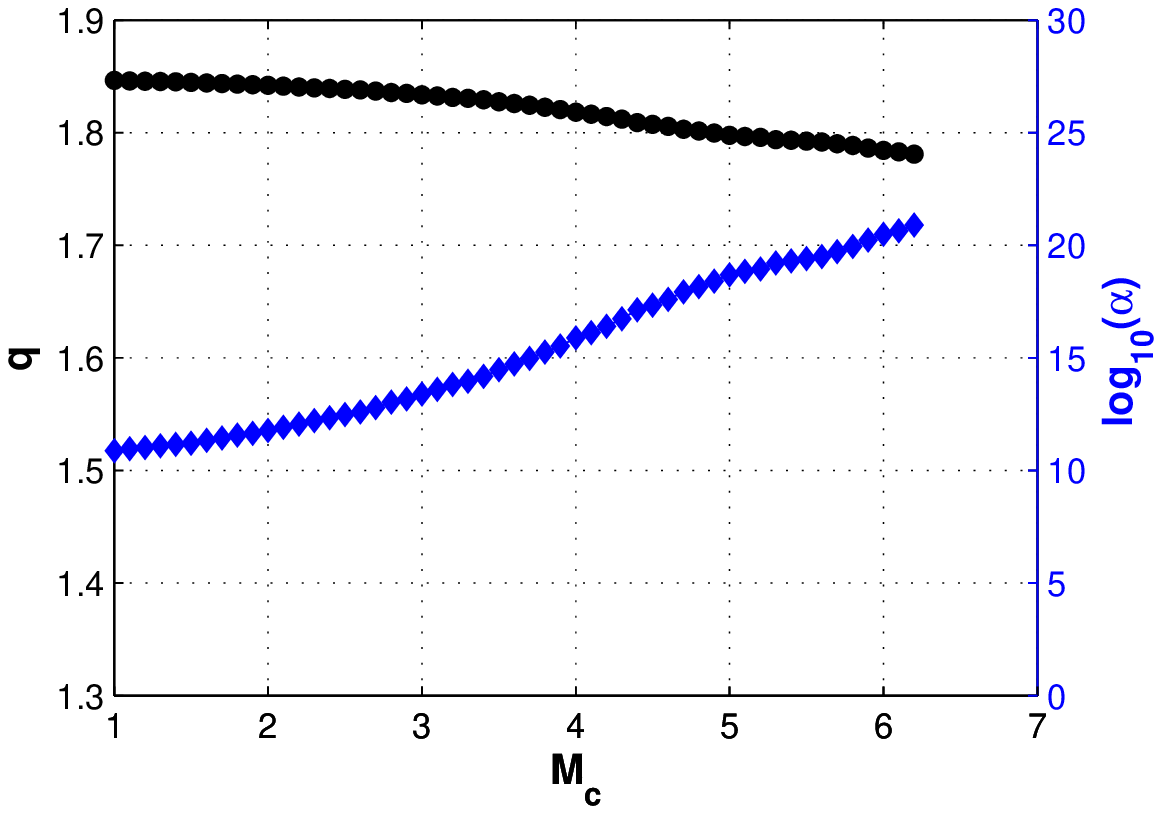}}           
\caption{Variation of nonextensive parameter $q$ and the volumetric energy density $\alpha $, for different thresholds of magnitudes of the detected EM-EQs for epoch 1 and epoch 2 respectively.}
\label{fig:tempvar}
\end{figure}

We recall that the scope of this analysis is to examine whether the two epochs of the precursory activity are really rooted in different regimes. The latter results support this hypothesis. In the following, we focus on the variation of the nonextensive parameter $q$ and the volumetric energy density $\alpha$ for different cutoffs of magnitudes, $M_c$, of the detected EM-EQs, included in the two epochs of the emerged EM precursor, as the EQ is approaching. For statistically valid results, we used different $M_c$ with an increment step of  $M_c=0.1$  and a minimum event number of $50$ events. For each step we fitted Eq. \ref{eq:Silva} to the available data points using the aforementioned LM method. In Figs. \ref{fig:tempvar1} and \ref{fig:tempvar2} the black curves depict the variation of the $q$-parameter for different $M_c$.

The observed high $q$-values for different threshold magnitudes reveal that the nonextensivity in the underlying fracture mechanism remains high in both epochs. This evidence is consistent with the hypothesis that the recorded EM precursor is associated with the final stage of EQ generation. Additionally, the prospective decrement of the nonextensive parameter $q$ as the magnitude threshold increases is explained by the fact that the larger the magnitude threshold the larger the number of EM-EQs/fractures is omitted. Indeed, the large number of the small fractures along with the corresponding redistribution of stresses significantly contributes to the increment of the correlation length during the fracture process \cite[and references therein]{Sornette2002}. The characteristic value that governs the overall system is the one that corresponds to the smaller magnitude threshold for each period. It should also be noted that although the nonextensive parameter $q$ decreases at higher magnitude thresholds it still remains high. This evidence further verifies that strong correlations have developed within the system. 

We recall that $\alpha$ is the coefficient of proportionality between fragment size and released energy in the fragment-asperity model \cite{Silva2006}. The blue curves in figs. \ref{fig:tempvar1} and \ref{fig:tempvar2} show that the energy density $\alpha$ characteristic value, increases at higher threshold values. This feature is consistent with the hypothesis that larger EM-EQs are rooted in larger and stronger entities. Special attention should be given to the following points of differentiation between the two highly non-nonextensive epochs 1 and 2: 

\begin{enumerate}[(i)]
\item {The maximum magnitude of EM-EQs in epoch 1 is $M_{max}=8.19$ while in epoch 2 it is $M_{max}=10.36$.}
\item {EM-EQs in the abruptly emerged epoch 2 reach significantly higher values of energy density $a$ in the order of $10^{21}$ while in epoch 1 reach values in the order of $10^{17}$.}
\end{enumerate}

\noindent Summarizing, up to this stage of the analysis we have shown that:  

\begin{enumerate}[(i)]
\item {The degree of the organization of the two strong EM bursts included in epoch 2 is significantly higher than in epoch 1.}
\item {The abrupt emerged two kHz EM bursts contained in epoch 2 are characterized by persistency in contrast to epoch 1 which is characterized by antipersistency.}
\item {Both two epochs are characterized by strong nonextensivity}.
\item {The second epoch is rooted in the fracture of larger and stronger entities in comparison to the first epoch}.
\end{enumerate}

The above mentioned characteristics imply that the two epochs of preseismic kHz EM activity under study refer to different models of EQ dynamics. In the following we attempt to establish the hypothesis that the first and second epochs are consistent with the nonextensive fragment and self-affine model, correspondingly.

\section{Footprints of persistent-fBm profile of fracture surfaces in the second epoch of kHz EM activity}
\label{sec:footprints}

Despite the complexity of the Earth's crust there are several universally holding scaling relations \cite[and references therein]{Eftaxias2009}. Such universal structural patterns of fracture and faulting process should be included into an associated EM precursor. Notice that Maslov et al. (1994) \cite{Maslov1994} have formally established the relationship between spatial fractal behavior and long-range temporal correlations for a broad range of critical phenomena. They showed that both the temporal and spatial activity can be described as different cuts in the same underlying fractal.

From the early work of Mandelbrot (1982) \cite{Mandelbrot1982}, much effort has been put to statistically characterise the resulting fractal surfaces in fracture processes:

\begin{enumerate}
\item {Fracture surfaces were found to be self-affine following the fractional Brownian motion (fBm) model over a wide range of length scales.}
\item {The spatial roughness of fracture surfaces has been interpreted as a universal indicator of surface fracture, weakly dependent on the nature of the material and on the failure mode.}
\end{enumerate}

Therefore, a fracture surface follows the persistent fBm-model, and consequently, an associated EM precursor should behave as a persistent fBm temporal fractal.

{\it Herein, we note that the characteristic of the persistent fBm model does not refer to a population of residual-fragments coming from previous fractures. Consequently, if our hypothesis is true, this characteristic should be contained only in epoch 2 due to the interference of the large and strong teeth.} 

\subsection{Footprints of universal roughness of fracture surfaces}

The Hurst Exponent $H$ specifies the strength of the irregularity (``roughness'') of the surface topography \cite{Turcotte1997} and it has also been interpreted as a universal indicator of fracture surfaces \cite{Lopez1998,Hansen2003,Zapperi2005,Mourot2006}. The height-height correlation function of a surface $\Delta h(r)=<[h(r+\Delta r)-h(r)]_{r}^{1/2}>$ computed along a given direction has been found to scale as $\Delta h\sim {{(\Delta r)}^{H}}$, where $H$ refers to the Hurst exponent. The Hurst exponent $H \sim 0.7$ has been interpreted as a universal indicator of surface fracture, weakly dependent on the nature of the material and the failure mode \cite{Lopez1998,Hansen2003,Ponson2006,Mourot2006,Zapperi2005}. Importantly, Renard et al., (2006) \cite{Renard2006} measured the surface roughness of a recently exhumed strike-slip fault plane by three independent 3D portable laser scanners. Their statistical scaling analyses revealed that the striated fault surface exhibits self-affine scaling invariance that can be described by a scaling roughness exponent, $H_l=0.7$ in the direction of slip. In Section \ref{sec:RSAnalysis} we showed that the ``roughness'' of the profile of the two strong kHz EM bursts, as it is represented by the Hurst exponent, is distributed around the value of 0.7. Thus, the universal spatial roughness of fracture surfaces nicely coincides with the roughness of the temporal profile of the recorded two strong preseismic kHz EM bursts included in the second epoch and does not coincide with the temporal profile of the preseismic time series of the first epoch. Herein, we recall the mean Hurst exponent of epoch 1 is $\bar{H}=0.38$.

\subsection{Footprints of fractional-Brownian-motion model of fracture surfaces}

If a pre-seismic EM time series behaves as a persistent temporal fractal, then, a power-law of the form $S(f) \propto f^{-\beta}$ is obeyed, with $S(f)$ the power spectral density (PSD) and $f$ the frequency. In a $\log S(f) - \log f$ representation the power spectrum is a line with slope $\beta$. The spectral scaling exponent $\beta$ is a measure of the strength of time correlations, while the quality of the fit of a time series to the power-law is represented by the linear correlation coefficient $r$. 

\begin{figure}[h]
  \centering
	\subfloat[Epoch 2(A)]{\label{subfig:bdist1}\includegraphics[width=0.4\textwidth]{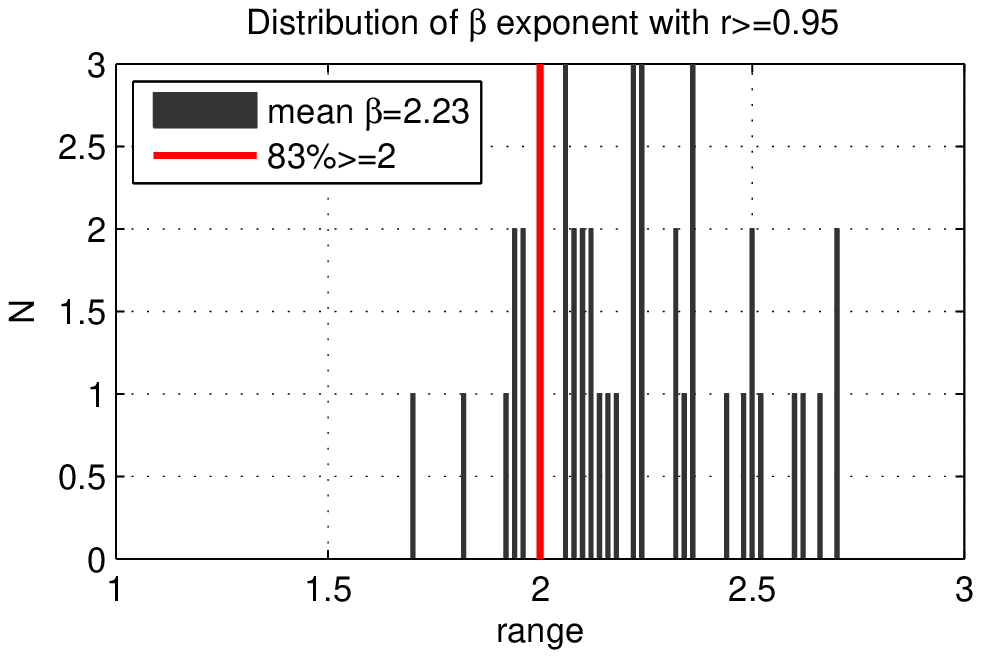}}     
  	\subfloat[Epoch 2(B)]{\label{subfig:bdist2}\includegraphics[width=0.4\textwidth]{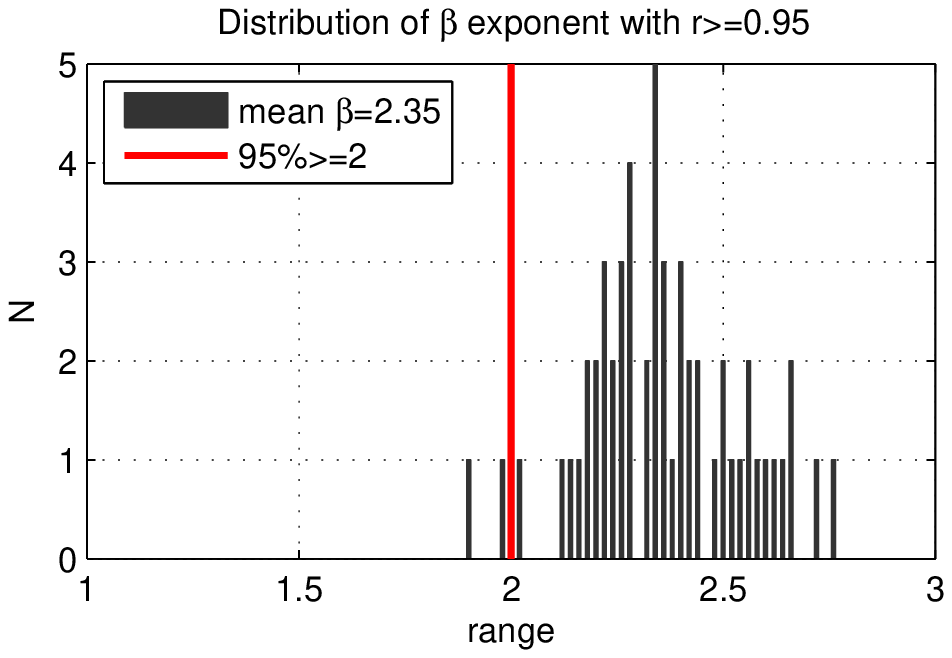}}       
\caption{The distribution of the $\beta$ - exponents with $r\ge 0.95$ as applied to epoch 2(A) and 2(B) respectively}
\label{fig:bdist}
\end{figure}

The "wavelet spectrum" is used in order to provide an unbiased and consistent estimation of the true power spectrum of the time-series with the 'Morlet' wavelet as a mother function. A fixed 1024 moving window with no overlapping sequence was used for the bursts, 2(A) and 2(B), see Fig. \ref{fig:athsig} respectively. For each window the local parameters $\beta$ and $r$ were derived. In Figs. \ref{subfig:bdist1} and \ref{subfig:bdist2}, we present the distribution of the $\beta$- exponents with $r\ge 0.95$. 

\begin{figure}[h]
  \centering
	\subfloat[Epoch 2(A)]{\label{subfig:HFit1}\includegraphics[width=0.4\textwidth]{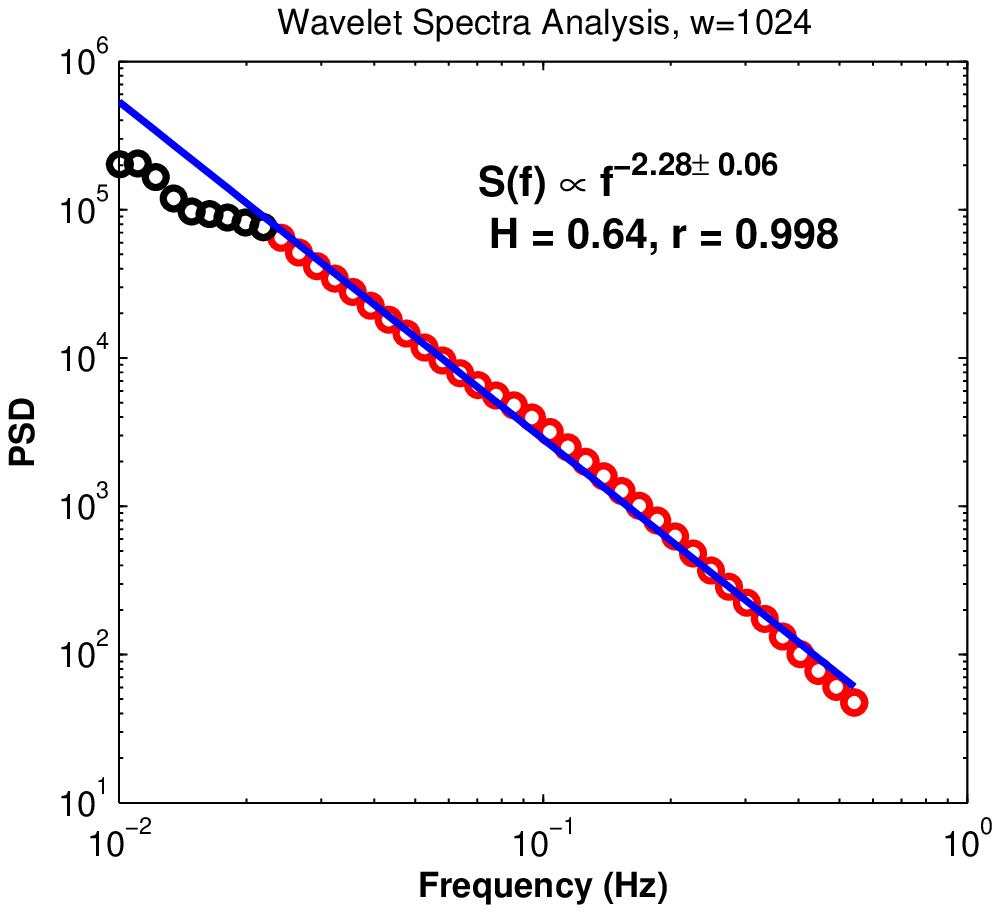}}     
  	\subfloat[Epoch 2(B)]{\label{subfig:HFit2}\includegraphics[width=0.4\textwidth]{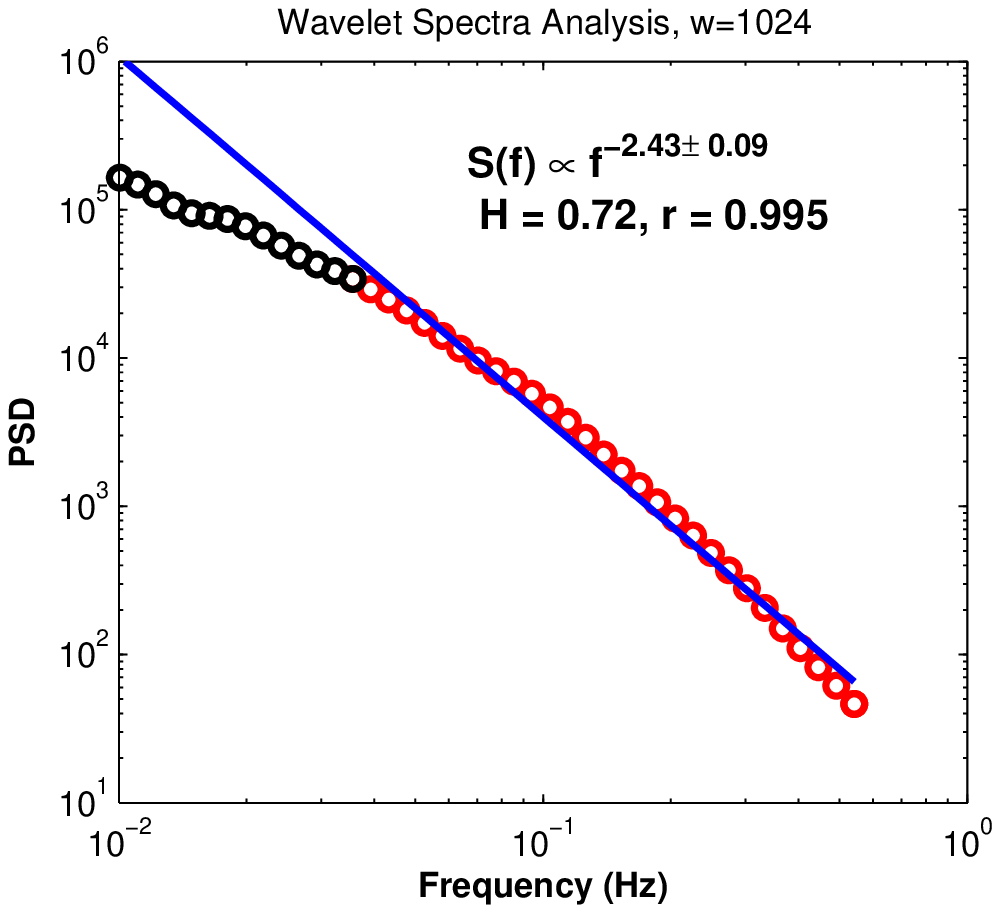}}           
\caption{Overall linear regression fitting for the estimation of the $\beta$ exponent of the power spectra density function: $S(f) \propto f^{-\beta}$}
\label{fig:HFit}
\end{figure}

Two classes of signal have been widely used to model stochastic fractal time series: fractional Gaussian noise (fGn) and fractional Brownian motion (fBm) \cite{Heneghan2000}. For the case of the fGn model the scaling exponent $\beta$ lies between $-1<\beta<1$, while the regime of fBm is indicated by $\beta$ values from 1 to 3. The distribution of $\beta $ - exponents reveal that the profile of the two strong EM bursts follows the fBm-model. To further support this finding, Figs. \ref{subfig:HFit1} and \ref{subfig:HFit2} show the linear regression fitting for the estimation of the $\beta$ exponents of the overall two strong EM bursts with values: $\beta =2.28 \pm 0.06$ and $\beta =2.43 \pm 0.09$ for the epochs 2(A) and 2(B). This finding verifies that the profile of the second epoch of the candidate kHz EM precursor is qualitatively analogous to the fBm-model ($1<\beta<3$), and could be originated during the slipping of two rough and rigid Brownian profiles \cite{De-Rubeis1996} that follow the fBm-model. 

We recall that the $\beta$ exponent is related to the Hurst exponent, $H$, by the formula $\beta = 2H +1$ with $0<H<1$ ($1<\beta<3$) for the fBm model \cite{Heneghan2000}. Fig. \ref{fig:HFit} shows the estimated values of Hurst exponent, $H=0.64$ and $H=0.72$, calculated by applying the $\beta$ - exponent to the formula: $\beta = 2H +1$. This finding further verifies that two strong EM bursts follow the persistent fBm-model. 

\begin{figure}[h]
  \centering
	\subfloat[Epoch 2(A)]{\label{subfig:HvsRS2}\includegraphics[width=0.33\textwidth]{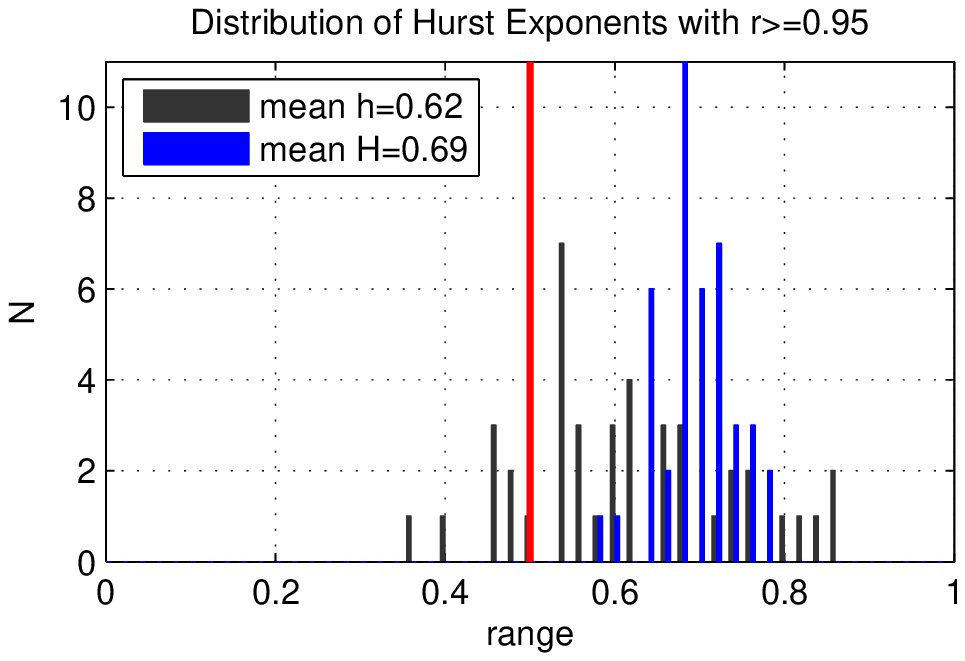}}    
   \subfloat[Epoch 2(B)]{\label{subfig:HvsRS3}\includegraphics[width=0.33\textwidth]{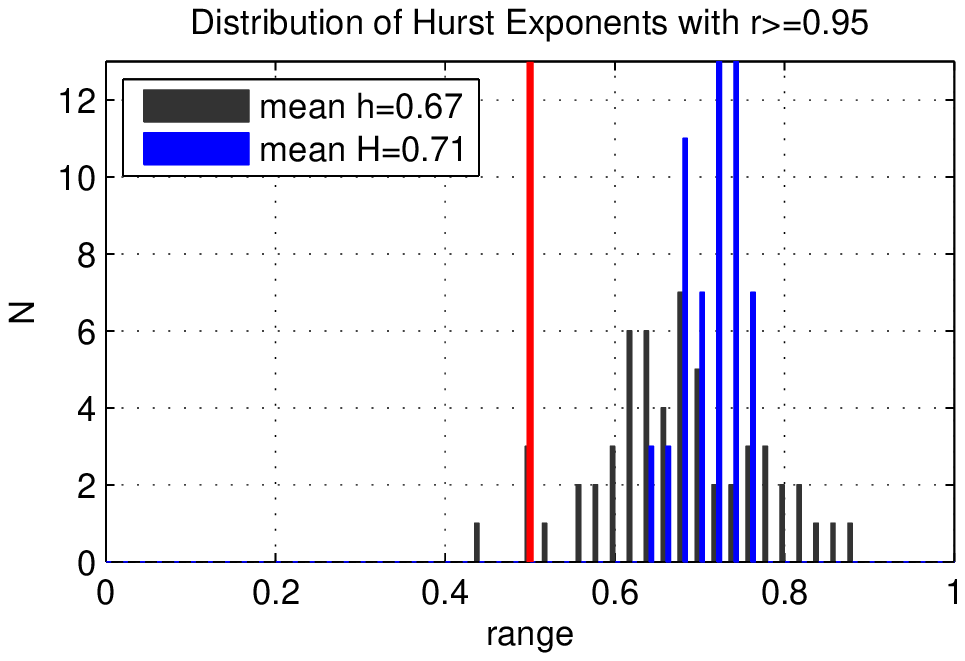}}    
	\subfloat[Epoch 1]{\label{subfig:HvsRS1}\includegraphics[width=0.33\textwidth]{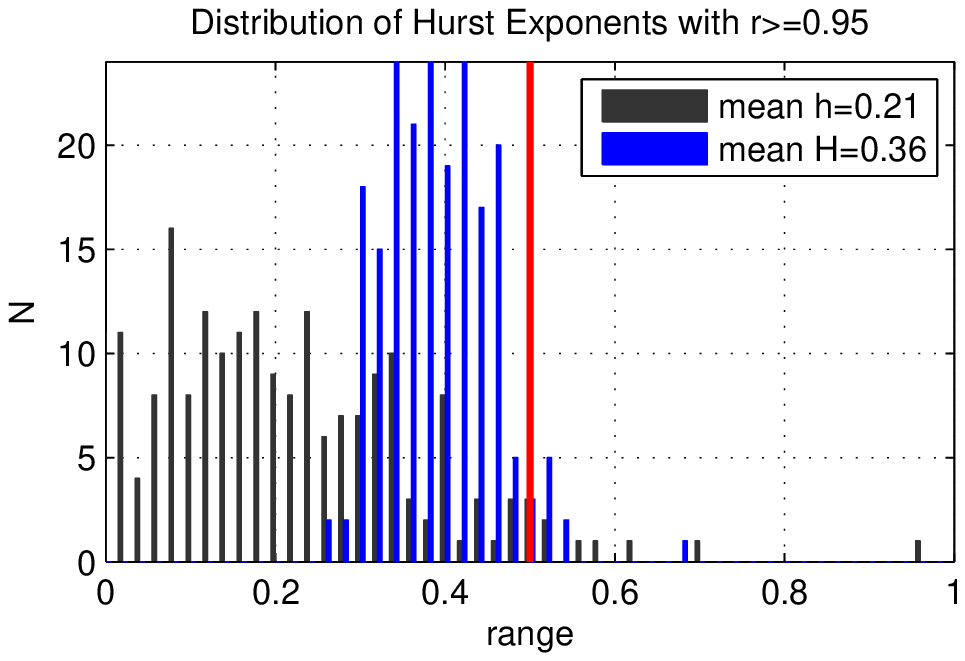}}     

\caption{The distribution of $H$-exponents deriving from the formula $\beta = 2H+1$ in relation to the distribution of Hurst $h$-exponents as they have been estimated in terms of R/S analysis, respectively from epoch 2(A), 2(B) and epoch 1.}
\label{fig:HvsRS}
\end{figure}

Fig. \ref{fig:HvsRS} depicts the distribution of $H$-exponents deriving from the formula $\beta = 2H+1$ in relation to the distribution of $H$-exponents as they have been estimated in terms of R/S analysis, for the epochs 2(A), 2(B) and 1, respectively. We observe that the two distributions (Figs. \ref{subfig:HvsRS2} and \ref{subfig:HvsRS3}) referred to the two strong EM bursts contained in epoch 2 concur, supporting the persistent-fBm temporal profile of the second epoch. On the contrary the $H$-exponent distributions of epoch 1 (Fig. \ref{subfig:HvsRS1}), verify the antipersistent profile which is not consistent with the universal indicator of fracture surfaces ($H\sim 0.7$).

\subsection{Additional arguments supporting the association of the second epoch with the self-affine asperity model}
\label{sec:arguments}

In this section we refer to arguments which further support the association of the second epoch with the self-affine asperity model introduced by De Rubeis et al. \cite{De-Rubeis1996} and Hallgass et al. \cite{Hallgass1997} .

According to the self-affine asperity model, the distribution of areas of the asperities broken $A$, follows a power law $P(A)~\sim {{A}^{-\delta }}$, with an exponent $\delta $, which could be related to the Hurst exponent ($0<H<1$) that controls the roughness of the fault. The former relation is obtained by supposing that the area of the broken asperities scales with its linear extension $l$ as $A_{asp}\sim l^{1+H}$. The H-exponent associated with the second epoch of the kHz EM activity is distributed around the value $0.7$. It is reasonable to assume that the broken teeth scaled with its linear dimension $l$ as $A_{asp}\sim l^{1.75}$. Importantly, numerical studies performed by de Arcangelis et al. (1989) \cite{Arcangelis1989} indicate that the number of bonds that break, scales during the whole process of fracture as $1.7$. This consistency further supports the association of the second epoch with the self-affine asperity model.

The self-affine model also reproduces the Gutenberg-Richter law suggesting that a seismic event releases energy in the interval $[E, E+dE]$ with a probability $P(E)dE$, $P(E) \sim E^{-B}$, where $B = \alpha + 1$ and $\alpha = 1 - H/2$ with $\alpha \in [1/2,1]$. In the present case, the Hurst-exponent $H \sim 0.7$ leads to $\alpha \sim 0.65$. Thus, the fracture of asperities released EM energies following the distribution $P(E) \sim E^{-B}$, where $B \sim 1.65$. This value is in harmony with geophysical data. Indeed, the distribution of energies released at any EQ is described by the power-law, $P(E) \sim E^{-B}$, where $B \sim 1.4-1.6$ \cite{Gutenberg1954}.

\section{Discussion \& Conclusions}

EM emissions in a wide frequency spectrum ranging from kHz to MHz are produced by opening cracks, which can be considered as precursors of general fracture. According to a recently proposed two-stage model on preseismic EM activity, the MHz EM emission is thought to be due to the fracture of the highly heterogeneous system that surrounds the fault. The finally emerged kHz EM emission is rooted in the final stage of EQ generation, namely, the fracture of entities sustaining the system. In this work we have further examined and elucidated the link of the precursory kHz EM activity with the last stage of EQ generation building on two theoretical models for EQ dynamics. The first self-affine asperity model states that the EQ is due to the slipping of two rough and rigid Brownian profiles one over the other. In this scheme, an individual EQ occurs when there is an intersection between the two fractional Brownian profiles. The second model, which is rooted in a nonextensive Tsallis framework starting from first principles, consists of two rough profiles interacting via fragments filling the gap. In this nonextensive approach, the mechanism of triggering EQ is established through the interaction of the irregularities of the fault planes and the fragments between them. This paper has attempted to show that two models of EQ dynamics supplement each other, in a sense, and both are mirrored in the detected two qualitative different epochs of the preseismic kHz EM emission associated with the Athens EQ. We have argued that the initially emerged precursory activity (epoch 1) follows the nonextensive model and is due to the fracture of fragments filling the gap between the two rough planes of the activated fault. The finally emerged precursory activity containing the two abruptly emerged strong impulsive EM bursts, follows the self-affine asperity model.

More precisely, our results illustrate that both epochs are characterized by strong nonextensivity. This finding supports the association of two epochs with the final stage of EQ generation. However, significant differentiations between the two distinct epochs of the recorded kHz EM activity, verify the link with the two models of EQ dynamics: 
  
\begin{enumerate}[(i)]
\item {The degree of the organization of the two strong EM bursts included in epoch 2 is significantly higher than in the initial epoch 1}. 
\item {Importantly, the abrupt emerged two kHz EM bursts in the tail of the preseismic EM activity (epoch 2) are characterized by persistency in contrast to epoch 1 which is characterized by antipersistency}. 

Recently, Sammis and Sornette (2002) \cite{Sammis2002} presented the most important mechanisms for such positive feedback. Laboratory experiments by means of acoustic and EM emission show that the main rupture occurs after the appearance of persistent behaviour (\cite{Ponomarev1997,Alexeev1993a,Alexeev1993b}. Recently Lei et al. (2000, 2004) \cite{Lei2000,Lei2004} have studied how individual and coupled asperities fracture as well as the role of asperities in fault nucleation and as potential precursors prior to dynamic rupture. The self-excitation strength, which expresses the strength of the effect of excitation associated with the preceding event on succeeding events, or equivalently, the degree of positive feedback in the dynamics, reaches the maximum value of $\sim 1$ during the nucleation.

\item {The second epoch is rooted in the fracture of larger and stronger entities in comparison to the first epoch. The maximum magnitude of EM-EQs in epoch 1 is $M_{max}=8.19$ while in epoch 2 it is $M_{max}=10.36$.}

\item {EM-EQs in the abruptly emerged epoch 2 reach significantly higher values of energy density $a$ of the order of $10^{2}$ while in epoch 1 energy reaches values up to $10^{17}$. We further emphasise that the breakage of fragments filling the gap is easier than the breakage of teeth distributed along the rough fault planes \cite{Sotolongo2004}.}

\item {The kHz EM precursor of the second epoch includes crucial universal feature of fracture of surfaces; namely, it follows the persistent fBm-model with a roughness consistent with the universal roughness of fracture surfaces ($H\sim 0.7$). This universal footprint is not mirrored in the first epoch.}  
\end{enumerate}

The key differentiations mentioned above support the following hypothesis: the first epoch refers to the fracture of fragments intervening between the two anomalous surfaces of the fault that contributes to the hindering of their relative motion. Once the fracture of one fragment has occurred, there is a reformation of fragments followed by a redistribution of stresses. This process practically results to a relative displacement of the fault planes (fault slip). The next EM-EQ will emerge when a new fragment breaks under the impact of the increased tensions. This process is consistent with the antipersistent character of the first epoch. As the fragments are broken, the two rough planes of the fault approaches each other. The abruptly emerged second epoch of large EM-EQs is the reflection of collision and breakup of large and strong teeth of the irregular surfaces. Its persistent behaviour and high organization along with the corresponding higher magnitudes $M$ and energy density $\alpha$ values supports this hypothesis. 

A challenging issue in the field of EQ dynamics is whether the generation of an EQ can be adequately explained by the self-affine model described in section \ref{self_affine_model} or the nonextensive fragment-asperity model as described in section \ref{fragment_asp_model}. Confronted with such a question, we may not seek answers merely in the statistics of EQs. Both two models of EQ dynamics lead to a power-law distribution of magnitudes. On the one hand, the self-asperity model leads to the traditional power-law distribution of magnitudes, known as Gutenberg-Richter law. On the other hand, it has been shown \cite{Sarlis2010} that the fragment-asperity model as described by Eq. \ref{eq:Silva} also leads to a Gutenberg-Richter-type - law for large EQs by the following equation \cite{Sarlis2010}:

\begin{equation}
b = 2 \times \frac{2-q}{1-q}
\label{eq:sarlis}
\end{equation}

\noindent The fact that both models find quantitative expression through the latter equation is not an unexpected result. Indeed, some generic behaviour seems to be shared by fragmenting systems whatever their size, material, or typical interaction energy. For instance, as observed in a large variety of experiments \cite{Sator2010,Turcotte1986} and natural phenomena \cite{Turcotte1986, Sammis1986,Kaminski1998, Carpinteri2005}, the fragment size distribution frequently exhibits a power law behavior, the origin of which is still unknown. Several examples of power-law fragmentation are given in \cite{Scholz1989}. Considering that the released energy $\varepsilon$ is proportional to fragment size \cite{Silva2006}, it is reasonable that the magnitudes of EQs which are rooted in the fracture of the population of fragments filling the space between fault planes also follow a power-law distribution.

In the research prospect of discriminating whether a seismic shock is sourced in the fracture of fragments filling the gap between the rough profiles or in the fracture of teeth distributed across the fractional Brownian profiles, it would be more appropriate to focus on the generation of a single EQ and not on the statistics of a population of different EQs. The reported significant differentiations of the two epochs imply that the fracture induced kHz pre-seismic EM emissions seem to offer such a possibility. Our proposed approach cannot claim any universal applicability or objective truth. The experimental results presented in this paper justify the link between nonextensive fragment-asperity and self-affine asperity models for EQ dynamics with preseismic kHz EM activity. However, future research needs to be undertaken in order to test the applicability of such an approach in different data-sets and further verify the key features identifying with epoch 1 and epoch 2 respectively, as proposed herein.

\section{Acknowledgments}
The first author (G.M) would like to acknowledge research funding received from the Greek State Scholarships Foundation (IKY).

\end{document}